\documentclass[preprint]{sig-alternate}

\usepackage{multirow} 
\usepackage{url} 

\usepackage{algorithm}
\usepackage{algpseudocode}
\usepackage{pseudocode}
\begin{document}
%

\title{Correlation of Expert and Search Engine Rankings}
\numberofauthors{3}
\author{
\alignauthor
Michael L. Nelson\\
       \affaddr{Department of Computer Science}\\
       \affaddr{Old Dominion University}\\
       \affaddr{Norfolk, VA,  23529}\\
       \email{mln@cs.odu.edu}
\alignauthor
Martin Klein\\
       \affaddr{Department of Computer Science}\\
       \affaddr{Old Dominion University}\\
       \affaddr{Norfolk, VA,  23529}\\
       \email{mklein@cs.odu.edu}
\alignauthor
Manoranjan Magudamudi\\
       \affaddr{Department of Computer Science}\\
       \affaddr{Old Dominion University}\\
       \affaddr{Norfolk, VA,  23529}\\
       \email{maguds@gmail.com}
}
\maketitle
\begin{abstract}

In previous research it has been shown that link-based web page metrics
can be used to predict experts' assessment of quality.  We are interested
in a related question: do expert rankings of real-world entities correlate
with search engine rankings of corresponding web resources?  For example,
each year US News \& World Report publishes a list of (among others)
top 50 graduate business schools.  Does their expert ranking correlate
with the search engine ranking of the URLs of those business schools?
To answer this question we conducted 9 experiments using 8 expert rankings
on a range of academic, athletic, financial and popular culture topics.
We compared the expert rankings with the rankings in Google, Live Search
(formerly MSN) and Yahoo (with list lengths of 10, 25, and 50).  In 57 search
engine vs. expert comparisons, only 1 strong and 4 moderate correlations
were statistically significant.  In 42 inter-search engine comparisons,
only 2 strong and 4 moderate correlations were statistically significant.
The correlations appeared to decrease with the size of the lists: the
3 strong correlations were for lists of 10, the 8 moderate correlations
were for lists of 25, and no correlations were found for lists of 50.

\end{abstract}


\section{Introduction}

As a society, we enjoy lists, presumably compiled by ``experts'',
that rank items, events, people, places, etc.  At best, these lists
are informative and help convey notions of quality in a compact manner.  
At worst, these lists can be misleading, biased, or overly simplified.
Regardless, lists proclaiming the top 10, 25 or 50 of various resources
are a persistent part of our culture.  

At the same time, search engines now play a central role in society.
The ``big 3'' search engines (SEs) -- Google, Live (formerly MSN), and
Yahoo -- are the primary tool for discovering web resources for many
people.  Acquiring a high ranking in SEs is so important that an entire
discipline and economy of search engine optimizers (SEOs) has developed
to help people raise the ranking of their web pages.  Thus SEs move
from a simple navigation and discovery aid to powerful cultural force.
In some sense, if a web page does not appear in the first few pages of a
SE's results for a particular query, it is as if it does not exist at all.

Given the power that expert lists and SEs have, we are interested in
their intersection.  In particular, we want to know if expert rankings of
``real-world'' resources such as movies, popular songs, universities and
university departments, companies, professional tennis players and cities
in the United States correlate to the search engine rankings of their
corresponding web resources.  It was our intuition that highly ranked
real-world resources would be correspondingly highly-ranked in SEs.
To answer this question, we selected 8 expert lists and from this we
created 9 separate tests.  For each test, we mapped the movie, artist,
place, etc. to a single URL.  We then created a program that will create
an ordinal ranking of the URLs in a SE independent of any keyword query.
We then used Kendall's Tau ($\tau$) to test for statistically significant
(p $<$ 0.05), moderate (0.40 $<$ $\tau$ $\leq$ 0.60) or strong (0.60
$<$ $\tau$ $\leq$ 0.80) correlations between the expert rankings and SE
rankings, and the inter-SE rankings.  We found fewer correlations than
we expected, and we also discovered that the correlations decreased as
the size of the list increased.

The result is that although highly ranked pages are likely to be quality
pages (cf. \cite{amento:auth-qual}), we cannot be sure that quality
real-world resources (e.g., athletes, movies, universities) have highly
ranked web pages.  This has implications for digital libraries and and
other systems that build collections by using only search engine APIs
\cite{DBLP:conf/la-web/KraftS03} or use the APIs to augment focused
crawling techniques \cite{996384,1065455,habing:ecdl2004}.
Our findings also suggest there is future work in determining what
are the additional factors of quality that are missed by conventional
hyperlink derived metrics such as PageRank \cite{brin1998als} and its
many variations.


\section{Related Work}

Although we are unaware of previous work that measures the correlation
of expert rankings of ``real-world'' objects with their corresponding
web resources, the quality of web search results has been the subject
of many previous studies.

\subsection{Quality and Authority in the Web}

``Does `Authority' mean Quality?'' is the question Amento et
al. \cite{amento:auth-qual} asked when they evaluated the potential
of link- and content-based algorithms to identify high quality web
pages. Human experts rated web documents from the Yahoo directory
related to five popular topics by their quality.  Amento et al. found
a high correlation between the rankings of the human experts leading to
the conclusion that there is a common notion of quality.  By computing
link-based metrics as well as analyzing the link neighborhood of the
web pages from their dataset they were able to evaluate the performance
of machine ranking methods.  Here too they found a high correlation
between in-degree, Kleinberg's authority score \cite{324140} and PageRank.
They isolated the documents that the human experts rated with good quality
and evaluated the performance of algorithms on that list in terms of
precision at $5$ and at $10$. In-degree e.g., has a precision at $5$
of $0.76$ which means on average almost $4$ of the first $5$ documents
it returns would be rated good by the experts. In general they find
that in-degree, authority score and PageRank are all highly correlated
with rankings provided by experts.  Thus, web document quality can be
estimated with hyperlink based metrics.

Upstill, Craswell and Hawking \cite{UpstillCH03b} studied the PageRank
and indegree of URLs for Fortune 500 and Fortune Most Admired companies.
They found companies on those lists averaged 1 point more PageRank
(via the Google toolbar's self-reported 0-10 scale) than companies
on the list.  They also found that IT companies typically had higher
PageRank than non-IT companies. Similar to \cite{amento:auth-qual},
they found indegree highly correlated with PageRank.

Bharat and Mihaila \cite{bharat:experts} propose a ranking scheme
based on authority where the most authoritative pages get the highest
ranking. Their algorithm is based on a special set of ``expert documents''
which are defined as web pages about a certain topic with many links
to non-affiliated web pages on that topic.  Non-affiliated pages are
pages from different domains and with sufficiently different IP address.
These expert documents are not chosen manually but automatically picked as
long as they meet certain requirements (sufficient out-degree, etc). In
response to a user query the most relevant expert documents are isolated.
The proposed scheme locates relevant links within the expert documents and
follows them to identify target pages.  These pages are finally ranked
according to the number and relevance of expert documents pointing
to them and presented to the end user. Bharat and Mihaila evaluated
their algorithm against three commercial search engines and found that
it performs either just as good or in some cases even better than the
top search engine when it comes to locating the home page of a specific
topic. The same is true for discovering relevant pages to topic (where
many good pages exist).

Rieh \cite{rieh:judgment} conducted a study on user's judgment of
information quality and cognitive authority in the web by observing
the user's searching behavior. The idea was to understand the factors
that influence user's judgment of quality and authority in the web. In
her work information quality on an operational level is defined as
``the extend to which users think that the information is useful, good,
current and accurate''. Cognitive authority is ``the extend to which
users think that they can trust the information''.  Rieh found that users
do predictive judgment (before opening the page) and evaluative judgment
(after opening the page) when it comes to the choice what page and item on
a page to look at. If the evaluative judgment does not correlate with the
expectations made in the predictive judgment the user usually starts a new
page or goes back to a previous one.  If the two judgments match however
the user stays on the page and uses its information.  She also found in
her experiments that users identify the facets characterizing cognitive
authority in the web as: trustworthiness, reliability, scholarliness,
credibility, officialness and authoritativeness. However for the subjects
she conducted the study with authority was more important for some search
tasks than for others. Looking for medicine e.g., authority was a major
concern but did not affect the subjects much for the travel research task.

Rieh and Belkin \cite{rieh:understanding} conducted a similar study
about people's decision making in respect to information quality and
cognitive authority in the WWW. This study confirms the intuition that
users of the web assess information quality based on source credibility
and authority. Authority can be seen on a institutional level e.g.,
academic or governmental institutions and on a personal level e.g.,
professional experts.  Another interesting finding of this work is that
users believe that the web is less authoritative and also less credible
than other, more conventional information systems.   

\subsection{Quality as a Factor in Web Page Ranking} 

Cho et al. \cite{cho:quality} observe a ``rich-get-richer'' phenomenon
where popular pages tend to get even more popular since search
engines repeatedly return popular pages first. As other studies by Cho
\cite{cho:impact, 1083683} and Baeza-Yates \cite{baeza-yates:structure} have shown,
PageRank is significantly biased against new (and thus unpopular) pages which
makes it problematic for these pages to draw the user's attention even
if they are potentially of high quality.  That means the popularity of a
page can be much lower than its actual quality.  Cho et al. propose page
quality as an alternative ranking method. By defining quality of a web
page as the probability that a user likes the page when seeing it for
the first time the authors claim to be able to alleviate the drawbacks
of PageRank. With the intuition from PageRank that a user that likes
the page will link to it the algorithm is able to identify new and high
quality pages much faster than PageRank and thus shorten the time it
takes for them to get noticed.   

\subsection{Quality of Web Documents}

Lim et at. \cite{lim:quality} introduce two models to measure the quality
of articles from an online community like Wikipedia without interpreting
their content. In the basic model quality is derived from the authority of
the contributors of the article and the contributions from each of them
(in number of words). The peer review model extends the basic model by
a review aspect of the article content. It gives higher quality to words
that ``survive'' reviews.

An approach to automatically predict information quality is given by Tang
et al. \cite{tang:predicting}.  Analyzing news documents they observe an
association between users quality score and the occurrence and prevalence
of certain textual features like readability and grammar.

\section{Experiment Design}

The following sections details how the expert lists that were chosen,
explains how we chose URLs to correspond with the entries in the expert
lists, and discusses the searching and ranking algorithms and other
operational details.

\subsection{Choosing Expert Lists}

We chose a variety of topics (2 academic, 2 financial, 2
athletic and 2 popular culture) as well as choose expert rankings that
are well-known.  The accuracy, criteria or bias of these rankings may
be critiqued, but that is not the purpose of this investigation.  We simply
accept the rankings as given from the experts.  They include (please note
that the URLs are likely to change over time): 

\begin{enumerate}

\item \textit{ARWU} -- The top 50 North \& Latin American Universities
as determined by the 2007 Academic Ranking of World Universities
\footnote{\url{http://www.arwu.org/rank/2007/ARWU2007_TopAmer.htm}}.

\item \textit{ATP} -- The top 50 male tennis players (as of
2008-01-28) as ranked by the Association of Tennis Professionals
\footnote{\url{http://www.atptennis.com/3/en/rankings/entrysystem/default.asp}}.

\item \textit{Billboard} -- The top 50 popular music
songs as determined by Billboard Magazine (as of 2008-01-28)
\footnote{\url{http://www.billboard.com/bbcom/charts/chart_display.jsp?g=Singles&f=The+Billboard+Hot+100}}.
This list is determined by a combination of sales and radio airplay.  This
list contained duplicates (artists with more than 1 song simultaneously on
the chart).  Since only the top 50 can be accessed without registration,
this ranking produced lists of n=\{9,21,42\} when duplicates were removed. 

\item \textit{Fortune} -- The 2007 top 50 American
public corporations as measured by gross revenue.
This list is published annually by Fortune Magazine
\footnote{\url{http://money.cnn.com/galleries/2007/fortune/0704/gallery.500top50.fortune/}}.

\item \textit{IMDB} -- The top 250 movies as
voted on by users of the Internet Movie Database
(\url{www.imdb.com})\footnote{\url{http://www.imdb.com/chart/top}}.
We used only the top 50 of 250 movies.  We split this ranking into
two lists: one that used only \url{imdb.com} URLs, and the other
that used only \url{en.wikipedia.org} URLs for the same movie.
For example, the URL for the 1990 movie ``Goodfellas'' was
\url{http://www.imdb.com/title/tt0099685/} in the first list and
\url{http://en.wikipedia.org/wiki/Goodfellas} in the second list.

\item \textit{Money} -- The 2007 top 50 ``best
places to live'' in the United States as determined by Money Magazine
\footnote{\url{http://money.cnn.com/galleries/2007/moneymag/0707/gallery.BPTL_top_100.moneymag/}}.  The 2007 list was a departure from previous lists in that it only 
featured very small cities and towns (e.g., Milton, Massachusetts (population
27,500) instead of Boston, Massachusetts).  

\item \textit{US News} -- The 2007 top 50 graduate business school
programs as ranked by US News \& World Report \footnote{
\url{http://www.usnews.com/usnews/edu/grad/rankings/mba/brief/mbarank_brief.php}}.  

\item \textit{WTA} -- The top 50 female tennis players
as ranked by the Women's Tennis Association (as of
2008-01-29)\footnote{\url{http://www.sonyericssonwtatour.com/2/rankings/singles_numeric.asp}}.

\end{enumerate}

\subsection{Mapping Resources to URLs}
\label{matching}

After the expert lists have been chosen, we began the process of mapping
their real-world objects to single URLs.  For some lists (ARWU, Fortune,
US News) this was easily done because each real-world object has a
canonical URL.  For the IMDB lists, the URLs are not quite canonical,
but they do come from two extremely well-known web sites: \url{imdb.com}
and \url{wikipedia.org}.  For the other lists (ATP, Billboard, Money,
WTA), judgment calls were needed to determine the best URL.

The ARWU list was the easiest of all: each university had a unique URL
that we could agree was the canonical URL for the university (e.g.,
\url{www.harvard.edu} for Harvard University).  While there are many
URLs available on the web that discuss Harvard University, it is our
intuition that \url{www.harvard.edu} is the correct choice for representing
all aspects of the university (as opposed to individual departments or the
basketball team).  Similarly, it was straight forward picking canonical
URLs for the Fortune 500 companies, although when faced with multiple
URLs, we chose the most general or public URL (e.g., \url{www.aig.com}
over \url{www.aigcorporate.com}).  The business schools generally had
nicely structured URLs (e.g., \url{mba.tamu.edu}), but several had paths
in their URLs that prove to be a limitation in some SEs (e.g., Yahoo's
\texttt{site} operator does not distinguish \url{www.nd.edu/~mba/}
from \url{www.nd.edu} see section \ref{querying} below).  The IMDB URLs
were directly taken from the IMDB web page.  We used Google to locate
their Wikipedia links to generate the second IMDB list.

For the ATP and WTA lists, it was less direct.  Although many URLs
were easy to discern (e.g., \url{www.rogerfederer.com}), we could
not locate suitable home pages for 24 of the ATP members and 19 of
the WTA members.  In those situations we used a Wikipedia page (e.g.,
\url{en.wikipedia.org/wiki/Alona_Bondarenko}).  This is could be due to
some of the tennis players not having a large English-language fan base.

For Money Magazine's Best Places to Live list, we always chose the
``official'' government URL over commercial, real-estate or tourism
related pages.  This proved difficult in the case of Oleny, Maryland
(rank \#17).  Olney is an unincorporated area and a ``census-designated
place'' in the larger region of Montgomery County, Maryland.  We could
not find an obvious government web page for Olney, and did not want to
use a commercial page (\url{www.olneymd.com}).  We ended up using the web
page for Montgomery County, Maryland (\url{www.montgomerycountymd.gov}),
although a strong case can be made for former commercial page as well.

Mapping Billboard popular songs to URLs was the most problematic.
Instead of trying to pick a URL to correspond to a song, we chose
the home page of the artist that released the song.  As noted above,
several artists have more than one song on the Billboard list at one time,
resulting in less than 50 URLs (acquiring data for the songs ranked 51-100
required registration).  Furthermore, 9 of the songs were credited to
more than one artist.  For example, the number one song at the time of
writing is ``Flow'' and is listed as ``Flo Rida Featuring T-Pain''.
In these cases, we chose the home page for the artist listed first
(i.e., Flo Rida) and not the featured artist (i.e., T-Pain).  The popular
music artists also presented problems similar to the Olney, MD example
described above.  We used only ``official'' pages that appeared to be
maintained by the artists themselves.   We did not use unofficial ``fan''
pages, although anecdotally we know fan pages are often of high quality.
Even more challenging was that many artists had multiple candidate URLs:
their official page, and their official myspace.com page.  In all but one
cases we chose the official page over the artists' myspace.com pages;
the group ``Playaz Circle'' (song \#49 on the Billboard list) appeared
to only have a myspace.com page.

\subsection{Creating an Ordinal Ranking of URLs from SE Queries}

We developed a Perl program that takes a list of URLs and queries search
engines to determine their relative ordering of those URLs.  We do not
determine a search engine's absolute ranking for any particular URL.
That is, we do not compute: \\

$rank(URL_A)=0.92$ 

$rank(URL_B)=0.73$

$rank(URL_C)=0.42$

...\\

We also are not interested in estimating the PageRank (or related
metrics), independent of SEs, through link neighborhoods or other
means: the SEs are the subject of our study, not the web graph itself.
Instead, using a variation of strand sort (illustrated in section
\ref{examplesort}), we simply determine that a search engine ranks the
URLs in order: \\

$rank(URL_A) \ge$
$rank(URL_B) \ge$
$rank(URL_C) \ge$
...\\

Note that the ranks of both the experts and search engines are ordinal
variables, so generally: \\

$distance(URL_A, URL_B) \neq distance(URL_B, URL_C)$.  \\

We ran the program several times, but the results we report are from the
machine \url{tango.cs.odu.edu} (IP 128.82.4.75) on February 8, 2008.
The program queried the APIs of Google, Live and Yahoo.  Although it
has been shown that search engine APIs return different results than
the public (human) interfaces \cite{1255237} and possibly use
a smaller index, we chose to use the APIs instead of ``page-scraping''
the results to avoid being denied access by the search engines.

Although the SE APIs can be queried for backlinks or ranking metrics,
previous research has shown that these values are not always accurate,
perhaps intentionally so to prevent reverse engineering of SE ranking
algorithms \cite{1255237}.  Note that it is not our goal to compute
the interval value of a particular URL in a given SE, but rather just
to produce an ordinal ranking of URLs for a SE.  We treat the SEs
as a black box ranking system and do not try to reverse engineer its
hyperlink-based methods.

With the exception of the rankings of international professional tennis
players, all the expert lists and the SE APIs queried are biased toward
the English language and lists of interest to the United States.  We made
no attempt to query non-English language SEs.

Ideally, we could submit all 50 URLs to a SE in a single query and
record the resulting ordering.  However, each SE has query length
limtiations for both characters and terms (discussed below) and queries
that exceed these limitations are silently truncated.  We must issue
a series of overlapping queries to create an ordinal ranking of URLs
relative to a specific SE.  To this end, we used a variation of
strand sort\footnote{\url{http://en.wikipedia.org/wiki/Strand_sort}}.
Strand sort is a sorting algorithm that
uses multiple intermediate data structures to temporarily store a sorted
subset of the data. These structures are eventually gathered together to
sort the entire list of data. This behavior makes it part of the family
of distribution algorithms.

\subsubsection{Querying Search Engine APIs}
\label{querying}

In order to determine the SE ranking of the URLs we must form unbiased
queries.  We do that by using the \texttt{site:} query modifier which is
supported by all three search engines. It works as a filter by restricting
the results to websites in the given domain only.  We query for several
URLs simultaneously (specified by $q$) and thus combine the URLs and
the \texttt{site:} modifier with the boolean \texttt{OR} operator
(also supported by all three search engines).  This boolean operator
returns results that match either side of the query string divided by the
\texttt{OR}.  Since our queries consist of URLs only, each with the same
modifier and combined with the boolean operator and no keywords added,
all search results have theoretically an equal opportunity to be returned
as the top result and ``only'' the search engine's ranking is dictating
the ranking of the URLs now.  We verified these searches were commutative:
the order of the URLs in the queries did not change the final rankings.
As an example, the query for Google and Live for the first five business
schools in the US News ranking would be:

\begin{small}
\begin{verbatim}
site:http://www.hbs.edu/ OR
site:http://www.gsb.stanford.edu/ OR
site:http://mba.wharton.upenn.edu/ OR
site:http://mitsloan.mit.edu/mba OR
site:http://www.kellogg.northwestern.edu/
\end{verbatim}
\end{small}

There are restrictions to using the search engine's APIs. Google
allows only $1000$ queries per day and the query length must
not exceed $2048$ bytes and $10$ words.  Yahoo searches are done
slightly differently because their \texttt{site:} modifier requires
a different syntax. It does not allow URI schemes like \textit{http}
in the query when using the modifier. It also allows only domain
names without a specific path following the top level domain or
country code e.g. \url{site:mitsloan.mit.edu/} is legitimate but
\url{site:mitsloan.mit.edu/mba} is not. Thus the Yahoo form of the above
query is:

\begin{small}
\begin{verbatim}
site:www.hbs.edu/ OR site:www.gsb.stanford.edu/ OR
site:mba.wharton.upenn.edu/ OR site:mitsloan.mit.edu/ OR
site:www.kellogg.northwestern.edu/
\end{verbatim}
\end{small}

Besides the syntax Yahoo also limits the queries to $5000$
per day.  Due to Yahoo's \texttt{site:} modifier syntax we can
not include Wikipedia URLs in our comparison with the Yahoo
search engine because all Wikipedia URLs follow the pattern
\url{http://en.wikipedia.org/wiki/certain_object} where the path of the URL
would be dismissed and only the ranking of the English Wikipedia
site is compared to all other URLs, resulting in erroneously high
score for the URL.

\subsubsection{An Example Ordinal Ranking of URLs}
\label{examplesort}

We illustrate creating an ordinal ranking of URLs with
an example.  Assume an unsorted list $UL$ with eight URLs
$(G,E,B,A,C,H,F,D)$. The expected outcome in the sorted list $SL$ will
be ranked in lexicographical order and we chose $q=3$.  The first $q$
URLs $(G,E,B)$ are queried against the search engine an the result is
sorted $(B,E,G)$. The overlap URL (the $q^{th}$ element), let us call
it $OL$, is the URL $G$ since it is the result with the lowest rank in
this subset of URLs.  The other two URLs $(B,E)$ are stored in $SL$.

In the next iteration we pull the next $q-1$ elements
from $UL$ and together with $OL=(G)$ form a new query $(G,A,C)$ for the
search engine. The result is $(A,C,G)$ indicating that $A$ and $C$ can
be ranked anywhere higher than $OL$ and thus need to be merged with the
elements in $SL$. First we take $A$ and query it together with $(B,E)$
and get the result $(A,B,E)$. Since $SL$ contains just these three
elements we are assured we found the correct rank for $A$. We know that
$C$ was ranked lower than $A$ and thus only need to query $C$ together
with all elements from $SL$ ranked below $A$. Thus we query $(C,B,E)$
and receive the result $(B,C,E)$ which we can append to the top ranked
result $A$. $SL$ now consists of $(A,B,C,E)$. $G$ remains the $OL$
since it was still the lowest ranked element in the subset and will now
(in the third iteration) be queried together with the next $q-1$ elements
from $UL$. The query $(G,H,F)$ returns $(F,G,H)$ which means $H$ as the
lowest ranked URLs will become the new $OL$ and $F$ and $G$ need to be
merged with all elements of $SL$.  First we query $F$ together with the
first $q-1$ elements from $SL$ and get the result $(A,B,F)$. This may
not be the final position of $F$ yet since $SL$ contains more than three
elements. All we know at this stage is that $F$ is ranked below $A$ and
$B$.  Thus we need to also query $(F,C,E)$ and will get $(C,E,F)$. Now
all elements in $SL$ are checked against $F$ and it turns out $F$
is the last element and thus can be appended to $SL$ which now holds
the ranking $(A,B,C,E,F)$.  

As the second part of this third iteration
we need to find the final position of $G$. We again know its ranked
lower than $F$ and since $F$ is the last element of $SL$ we can simply
append $G$ to $SL$ which now contains the sorted list $(A,B,C,E,F,G)$.
The new $OL$ is queried together with the remaining element of $UL$,
$D$ and the query returns $(D,H)$. This result tells us we need to treat
$D$ the same way like we did with $F$ in the third iteration. We query
$(D,A,B)$ and get $(A,B,D)$ then we query $(D,C,E)$ and get the result
$(C,D,E)$. Now we have determined the final position of URL $D$ and can
place it accordingly in $SL$. Since the $OL$ is still $H$ and $UL$ is
empty we are assured $H$ is the lowest ranked URL in the entire set and
can simply append $H$ to $SL$. This is the final step of the algorithm
and $SL$ now holds the sorted list containing all URLs $(A,B,C,D,E,F,G)$.

\begin{small}
\begin{algorithm} 
\caption{Ranking Algorithm} \label{alg:ranking} 
\begin{algorithmic}[1] 
 \Procedure{Init}{Q} 
  \State let $Q$ be the list of all URLs to be sorted and
  let $q$ be the number of URLs compared at a time
  \State $FinalRankedList=undef()$
  \State take top $q$ URLs from $Q$ 
  \State issue the $q$ URLs to SEs and store ranked result set in $TmpRankedList$
  \State move top $q-1$ URLs from $TmpRankedList$ to $FinalRankedList$
  \State $OverlapURL=q^{th}$ ranked URL from $TmpRankedList$
  \While{$Q$ is not empty}
   \State take next top $q-1$ URLs from $Q$ 
   \State issue the $q-1$ URLs plus $OverlapURL$ to SEs and store ranked result set in $TmpRankedList$
   \State $TmpList=$ URLs ranked higher than $OverlapURL$ in $TmpRankedList$
   \If{$TmpList$ is empty} \Comment{($OverlapURL$ is the top result, thus $FinalRankedList$ is sorted)}
    \State move $OverlapURL$ to $FinalRankedList$
    \State add top $q-1$ URLs of $TmpRankedList$ to $FinalRankedList$
   \Else
    \State $FinalRankedList=Compare(TmpList,FinalRankedList)$
   \EndIf
   \State append $TmpRankedList$ to $FinalRankedList$
   \State $OverlapURL=q^{th}$ranked URL from $TmpRankedList$
  \EndWhile 
 \State \textbf{return} $FinalRankedList$
 \EndProcedure 
%
 \Procedure{Compare}{$TmpList,FinalRankedList$}
  \State $TmpFinalList=FinalRankedList$
  \State $WorkList=undef()$
  \ForAll{URL $i$ in $TmpList$} \Comment{(if $TmpFinalList$ is empty add the remaining URLs in $TmpList$ to $Worklist$)}
   \While{$TmpFinalList$ is not empty}
    \State take top $q-1$ URLs from $TmpFinalList$
    \State issue the $q-1$ URLs plus the $i^{th}$ URL to SEs and store ranked result set in $TmpRankedList$
    \If{$i$ is the last element in $TmpRankedList$}
     \State move the top $q-1$ URLs from $TmpRankedList$ to $WorkList$
    \Else
     \State move all URLs ranked higher than $i$ in $TmpRankedList$ to $WorkList$
     \State move $i$ to $WorkList$
     \State unshift the other URLs back to $TmpFinalList$ \Comment{(for comparing to the remaining URLs in $TmpList$)}
     \State break \Comment{(and move on to the next URL)}
    \EndIf
   \EndWhile
  \EndFor
  \State push all elements from $TmpFinalList$ to $WorkList$
  \State $FinalRankedList=WorkList$
  \State \textbf{return} $FinalRankedList$
 \EndProcedure 
\end{algorithmic} 
\end{algorithm}
\end{small}

\section{Results}
\subsection{Correlations}

Our null hypothesis ($H_0$) was that there is no correlation between
any of the rankings (experts and SEs as well as inter-SE).  Tables
\ref{tab:arwu} through \ref{tab:wta} show the Kendall's $\tau$ and
2-side p-value for each test.  Statistically significant (p $<$ 0.05)
moderate and strong correlations are bolded.  We compare each search
engine with the expert ranking as well as inter-search engine comparisons.

\begin{figure} [h!]
\begin{center}
\includegraphics[scale=0.33]{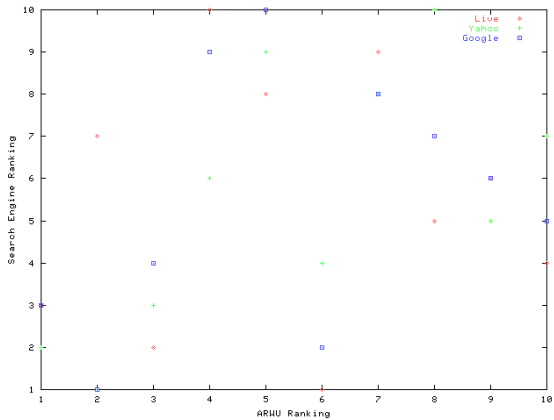}
\caption{ARWU and Search Engine Rankings (n=10).} 
\label{fig:arwu-10}
\end{center}
\end{figure}

\begin{figure} [h!]
\begin{center}
\includegraphics[scale=0.33]{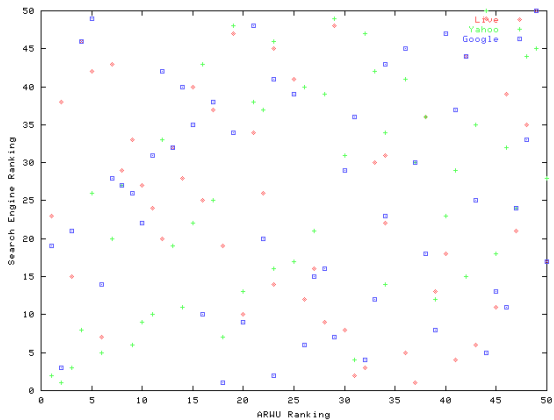}
\caption{ARWU and Search Engine Rankings (n=50; Key is the Same as Figure \ref{fig:arwu-10}).} 
\label{fig:arwu-50}
\end{center}
\end{figure}

We omit nearly all of the scatter plots because there is not enough of
a correlation for them to be useful.  For example, although there is a
moderate correlation between ARWU and Yahoo (n=10), looking at figure
\ref{fig:arwu-10} it is difficult to discern this correlation; the Yahoo
data appears very similar to the Live and Google data.  Furthermore,
when there is no correlation at all, like n=50 for the ARWU data, the
scatter plot is just noise (figure \ref{fig:arwu-50}).

We could only reject $H_0$ in 11 of 99 cases.  In 57 search engine
vs. expert comparisons, only 1 strong (table \ref{tab:wta} Live/WTA n=10)
and 4 moderate correlations (table \ref{tab:arwu} Yahoo/ARWU n=10 and
n=25, table \ref{tab:money} Google/Money n=10, table \ref{tab:usnews}
Google/US News n=25) were statistically significant.  Interestingly, the
Google/Money n=10 correlation was negative.  In 42 inter-search engine
comparisons, only 2 strong (table \ref{tab:arwu} Yahoo/Google n=10,
table \ref{tab:fortune} Live/Yahoo n=10) and 4 moderate correlations
(table \ref{tab:arwu} Live/Google n=10, table \ref{tab:fortune}
Live/Yahoo n=25, table \ref{tab:money} Live/Yahoo n=10 and n=25) were
statistically significant.  

The correlations appeared to decrease with the size of the lists: the 3
strong correlations were for lists of 10 and the 8 moderate correlations
were for lists of 25.  No correlations were found for lists of 50.
This is interestingly in contrast with \cite{melucci:rank-corr}, which
warns of $\tau$ increasing as the size of the lists grows.

\begin{table}
\begin{center}
\begin{tabular}{|c||c|l|l|}
\hline
Comparison & n & $\tau$ & p \\
\hline
\multirow{3} {*} {Live/ARWU} & 10 & -0.0222 & 1 \\
& 25 & 0.0066 & 0.9813 \\
& 50 & -0.1167 & 0.2349 \\
\hline
\multirow{3} {*} {Yahoo/ARWU} & \bf{10} & \bf{0.5111} & \bf{0.0490} \\
& \bf{25} & \bf{0.4666} & \bf{0.0011} \\
& 50 & 0.3436 & 0.0004 \\
\hline
\multirow{3} {*} {Google/ARWU} & 10 & 0.1555 & 0.5915 \\
& 25 & 0.0733 & 0.6238 \\
& 50 & 0.0008 & 1 \\
\hline
\multirow{3} {*} {Live/Yahoo} & 10 & 0.2000 & 0.4742 \\
& 25 &0.2599 & 0.0721 \\
& 50 & 0.1183 & 0.2283 \\
\hline
\multirow{3} {*} {Live/Google} & \bf{10} & \bf{0.5555} & \bf{0.0318} \\
& 25 & 0.2666 & 0.0650 \\
& 50 & 0.1151 & 0.2415 \\
\hline
\multirow{3} {*} {Yahoo/Google} & \bf{10} & \bf{0.6444} & \bf{0.0122} \\
& 25 & 0.2066 & 0.1542 \\
& 50 & -0.0775 & 0.4316 \\
\hline
\end{tabular}
\caption{\label{tab:arwu}SE and ARWU Ranking of North and Latin American Universities.}
\end{center}
\end{table}

\begin{table}
\begin{center}
\begin{tabular}{|c||c|l|l|}
\hline
Comparison & n & $\tau$ & p \\
\hline
\multirow{3} {*} {Google/ATP} & 10 & 0.1111 & 0.7205 \\
& 25 & 0.3933 & 0.0062 \\
& 50 & -0.09877 & 0.3154 \\
\hline
\end{tabular}
\caption{\label{tab:atp}SE and ATP Ranking of Male Tennis Players.}
\end{center}
\end{table}

\begin{table}
\begin{center}
\begin{tabular}{|c||c|l|l|}
\hline
Comparison & n & $\tau$ & p \\
\hline
\multirow{3} {*} {Live/Billboard} & 9 & 0.2777 & 0.3480 \\
& 21 & -0.0666 & 0.6946 \\
& 42 & -0.1045 & 0.3347 \\
\hline
\multirow{3} {*} {Yahoo/Billboard} & 9 & -0.0555 & 0.9169 \\
& 21 & 0.0666 & 0.6946 \\
& 42 & -0.1126 & 0.2981 \\
\hline
\multirow{3} {*} {Google/Billboard} & 9 & -0.3333 & 0.2514 \\
& 21 & -0.1428 & 0.3811 \\
& 42 & -0.1010 & 0.3513 \\
\hline
\multirow{3} {*} {Live/Yahoo} & 9 & -0.2222 & 0.4655 \\
& 21 & -0.1047 & 0.5259 \\
& 42 & 0.0894 & 0.4101 \\
\hline
\multirow{3} {*} {Live/Google} & 9 &0.1666 & 0.6021 \\
& 21 & 0.2761 & 0.0852 \\
& 42 & 0.2497 & 0.0203 \\
\hline
\multirow{3} {*} {Yahoo/Google} & 9 & -0.2777 & 0.3480 \\
& 21 & -0.0857 & 0.6077 \\
& 42 & -0.0987 & 0.36264 \\
\hline
\end{tabular}
\caption{\label{tab:billboard}SE and Billboard Ranking of Singles.}
\end{center}
\end{table}

\begin{table}
\begin{center}
\begin{tabular}{|c||c|l|l|}
\hline
Comparison & n & $\tau$ & p \\
\hline
\multirow{3} {*} {Live/Fortune} & 10 & -0.0222 & 1 \\  
& 25 & 0.1933 & 0.1831 \\
& 50 & -0.0612 & 0.5359 \\
\hline
\multirow{2} {*} {Yahoo/Fortune} & 10 & -0.0222 & 1 \\ 
& 25 & 0.2399 & 0.0972 \\
\hline
\multirow{3} {*} {Google/Fortune} & 10 & -0.2444 & 0.3710 \\ 
& 25 & 0.2066 & 0.1542 \\
& 50 & 0.0481 & 0.6275 \\
\hline
\multirow{2} {*} {Live/Yahoo} & \bf{10} & \bf{0.7333} & \bf{0.0042} \\
& \bf{25} & \bf{0.5133} & \bf{0.0003} \\
\hline
\multirow{3} {*} {Live/Google} & 10 & 0.4222 & 0.1074 \\
& 25 & 0.3866 & 0.0072 \\
& 50 & 0.3877 & 0.0001  \\
\hline
\multirow{2} {*} {Yahoo/Google} & 10 & 0.3333 & 0.2104 \\
& 25 & 0.4199 & 0.0035 \\
\hline
\end{tabular}
\caption{\label{tab:fortune}SE and Fortune Magazine Ranking of Companies.}
\end{center}
\end{table}

\begin{table}
\begin{center}
\begin{tabular}{|c||c|l|l|}
\hline
Comparison & n & $\tau$ & p \\
\hline
\multirow{3} {*} {Live/IMDB} & 10 & -0.2888 & 0.2831 \\
& 25 & 0.1799 & 0.2157 \\
& 50 & 0.2702 & 0.0057 \\
\hline
\multirow{3} {*} {Google/IMDB} & 10 & -0.2000 & 0.4742 \\
& 25 & 0.0999 & 0.4982 \\
& 50 & 0.0253 & 0.8018 \\
\hline
\multirow{3} {*} {Live/Google} & 10 & 0.2000 & 0.4742 \\
& 25 & 0.1066 & 0.4690 \\
& 50 & -0.0775 & 0.4316 \\
\hline
\end{tabular}
\caption{\label{tab:imdb}SE and IMDB Ranking of Movies.}
\end{center}
\end{table}

\begin{table}
\begin{center}
\begin{tabular}{|c||c|l|l|}
\hline
Comparison & n & $\tau$ & p \\
\hline
\multirow{3} {*} {Google/IMDB-wiki} & 10 & 0.4666 & 0.0736 \\
& 25 & 0.0799 & 0.5911 \\
& 50 & 0.0824 & 0.4028 \\
\hline
\end{tabular}
\caption{\label{tab:imdb-wiki}SE and IMDB Ranking of Movies (Wikipedia URLs).}
\end{center}
\end{table}

\begin{table}
\begin{center}
\begin{tabular}{|c||c|l|l|}
\hline
Comparison & n & $\tau$ & p \\
\hline
\multirow{3} {*} {Live/Money} & 10 & 0.0666 & 0.8580 \\
& 25 & 0.1933 & 0.1831 \\
& 50 & 0.0432 & 0.6635 \\
\hline
\multirow{3} {*} {Yahoo/Money} & 10 & 0.2444 & 0.3710 \\
& 25 & 0.1866 & 0.1989 \\
& 50 & 0.0726  & 0.4616 \\
\hline
\multirow{3} {*} {Google/Money} & \bf{10} & \bf{-0.5111} & \bf{0.04909} \\
& 25 & -0.0866 & 0.5593 \\
& 50 & -0.0987 & 0.3154 \\
\hline
\multirow{3} {*} {Live/Yahoo} & \bf{10} & \bf{0.5555} & \bf{0.0318} \\
& \bf{25} & \bf{0.5266} & \bf{0.0002} \\
& 50 & 0.3665 & 0.0001 \\
\hline
\multirow{3} {*} {Live/Google} & 10 & 0.0666 & 0.8580 \\
& 25 & -0.0399 & 0.7972 \\
& 50 & -0.0008 & 1 \\
\hline
\multirow{3} {*} {Yahoo/Google} & 10 & -0.3777 & 0.1524 \\
& 25 & -0.2733 & 0.0585 \\
& 50 & -0.2097 & 0.0322 \\
\hline
\end{tabular}
\caption{\label{tab:money}SE and Money Magazine Ranking of Places to Live.}
\end{center}
\end{table}

\begin{table}
\begin{center}
\begin{tabular}{|c||c|l|l|}
\hline
Comparison & n & $\tau$ & p \\
\hline
\multirow{3} {*} {Live/US News} & 10 & -0.2888 & 0.2831 \\
& 25 & 0  & 1 \\
& 50 & 0.1510 & 0.1237 \\
\hline
\multirow{3} {*} {Google/US News} & 10 & 0.2444 & 0.3710 \\
& \bf{25} & \bf{0.4199} & \bf{0.0035} \\
& 50 & 0.3142 & 0.0013 \\
\hline
\multirow{3} {*} {Live/Google} & 10 & 0.4666 & 0.0736 \\
& 25 & 0.1533 & 0.2932 \\
& 50 & 0.3240 & 0.0009 \\
\hline
\end{tabular}
\caption{\label{tab:usnews}SE and US News Ranking of Business Schools.}
\end{center}
\end{table}

\begin{table}
\begin{center}
\begin{tabular}{|c||c|l|l|}
\hline
Comparison & n & $\tau$ & p \\
\hline
\multirow{1} {*} {Live/WTA} & \bf{10} & \bf{0.6444} & \bf{0.0122} \\
\hline
\multirow{3} {*} {Google/WTA} & 10 & -0.1555 & 0.5915 \\ 
& 25 & 0.0634 & 0.6741 \\
& 50 & 0.1796 & 0.0669 \\
\hline
\multirow{1} {*} {Live/Google} & 10 & -0.1555 & 0.5915 \\ 
\hline
\end{tabular}
\caption{\label{tab:wta}SE and WTA Ranking of Female Tennis Players.}
\end{center}
\end{table}

\subsection{SE Errors}

Of the 9 tests, we were able to complete only 3 in all configurations:
for 3 list (n) sizes, 3 expert-SE comparisons and 3 inter-SE
comparisons.  These were ARWU (table \ref{tab:arwu}), Billboard (table
\ref{tab:billboard}), and Money (table \ref{tab:money}).  

Limitations of the Yahoo \url{site} operator (see section \ref{querying})
limited Yahoo's inclusion in ATP (table \ref{tab:atp}), both IMDB
tests (tables \ref{tab:imdb} and \ref{tab:imdb-wiki}), US News (table
\ref{tab:usnews}), and WTA (table \ref{tab:wta}).  There was a transient
error with Yahoo in the Fortune list for n=50 (table \ref{tab:fortune})
that we were unable to resolve on the day of the tests (15 URLs came
back as not indexed).  This problem disappeared on later runs, but rather
than report data for the Fortune list for a date other than February 8,
2008, we simply dropped the Fortune n=50 data.  This kind of transient
error in using SE APIs is consistent with the experiences reported in
\cite{1255237}.

Live produced unexpected results for all Wikipedia URLs.  We have no
explanation for why this is so, but it did result in Live being excluded
from ATP (table \ref{tab:atp}), IMDB-Wiki (table \ref{tab:imdb-wiki}),
and WTA n=25 and n=50 (table \ref{tab:wta}).  Interestingly, WTA n=10
did include 1 Wikipedia URL (which is returned as not indexed), but
Live still showed a strong correlation.  There were too many Wikipedia
URLs in n=25 and n=50 for the data to be meaningful.  Although we
queried the API, this behavior was seen in the human interface as well.
For example, Live indexes \url{en.wikipedia.org/wiki/Pulp_Fiction}.
But a query for \url{site:http://en.wikipedia.org/wiki/Pulp_Fiction}
returned did not return results about the movie, but instead
provided only the location of a local movie theater.  Also, queries
for \url{site:http://en.wikipedia.org/wiki/The_Godfather}
would produce only a single hit, 
\url{http://en.wikipedia.org/wiki/The_Godfather/Sandbox},
an invalid URL.

\begin{figure}[h!]
\begin{center}
\includegraphics[scale=0.30]{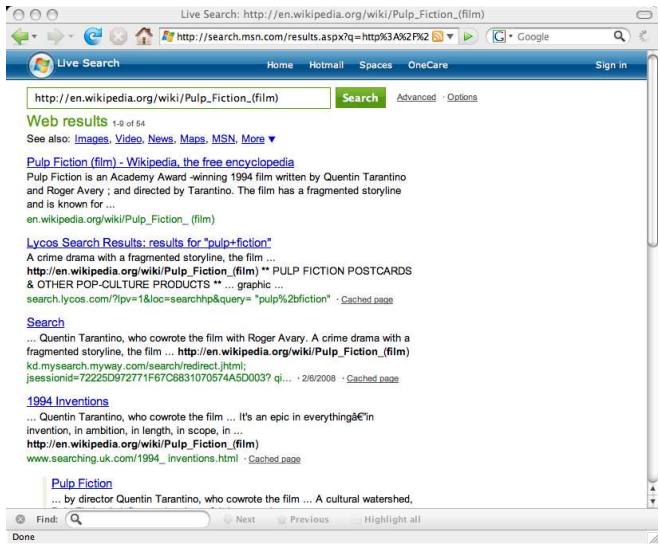}
\caption{Live Correctly Indexes the URL}
\label{fig:pulp1}
\end{center}
\end{figure}

\begin{figure}[h!]
\begin{center}
\includegraphics[scale=0.30]{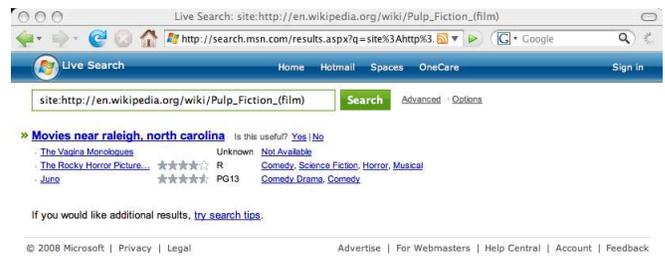}
\caption{Live Returns Incorrect Results for with the site operator}
\label{fig:pulp2}
\end{center}
\end{figure}

\begin{figure}[h!]
\begin{center}
\includegraphics[scale=0.28]{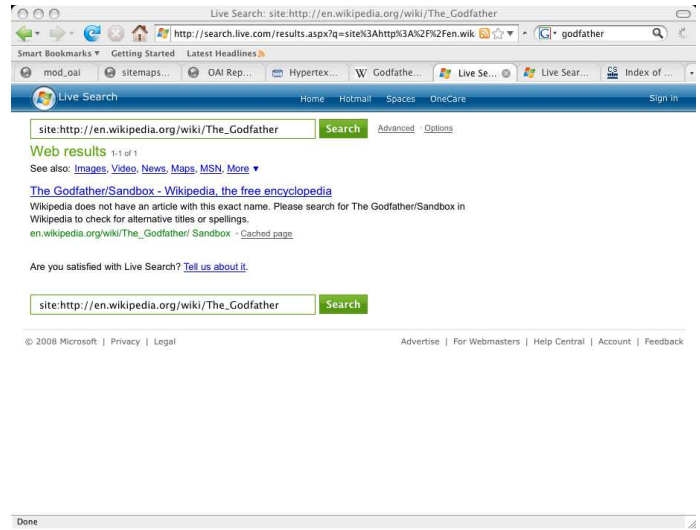}
\caption{A Different Type of Incorrect Result from Live with the site operator}
\label{fig:godfather}
\end{center}
\end{figure}

For example, figure \ref{fig:pulp1} shows that Live
indexes \url{en.wikipedia.org/wiki/Pulp_Fiction}.  But a query for
\url{site:http://en.wikipedia.org/wiki/Pulp_Fiction} shows that Live
does not return the hits shown in figure \ref{fig:pulp1}, but is trying
to provide location of a local movie theater (figure \ref{fig:pulp2}).
Figure \ref{fig:godfather} shows a clearly incorrect result when searching
for \url{site:http://en.wikipedia.org/wiki/The_Godfather}.

\section{Discussion}
\label{discussion}

We found fewer correlations than we anticipated.  We expected to find more
of both expert-SE correlations and inter-SE correlations.  The latter is
especially surprising; we are aware that crawling and ranking strategies
differ among the SEs, but they all are observing the same web graph (or at
least have the opportunity to observe the same web graph).  Although it is
well known that SE results have little overlap (e.g., \cite{spink2006sro,
bharat1998tmr, gulli:indexable}), we were not interested in, for example,
the rank of ATP players for the query ``tennis''.  Instead, we are
interested only in the ordinal ranking of ATP players in a given SE.

One possible reason that we did not observe more correlations
is that the methodology for mapping real-world objects to URLs
(section \ref{matching}) is limited.  For some lists, not every
entry has a clear canonical URL, and when forced to make a choice,
we may have chosen poorly (although their is no such problem with
the ARWU, Fortune and US News lists).  The mapping of Billboard
popular songs to the home pages of their respective artists is the
most tenuous.  Perhaps popular music lists of artists or their albums
would have performed better.  However, we did like the immediacy of
the Billboard list, and we know that some SEs will boost the ranking
of certain pages for current events and ``hot topics'' (cf. Google
Trends\footnote{\url{http://www.google.com/trends/hottrends}}).

It is possible that we have rediscovered the timeless discrepancy
of ``experts'' vs. ``popularity''.  The SE rankings are constantly
evolving to stay ahead of the spammers, but we believe they all
have some form of hyperlink-based popularity metric at their core
(i.e., variations on PageRank).  In anticipation of this, we did try
to include expert rankings that are closely aligned with general
notions of popularity.  For example, Billboard is determined by
sales and radio airplay.  The IMDB list is determined by popular
vote of IMDB users (cf. the American Film Institute (AFI) Top 100
movies\footnote{\url{http://connect.afi.com/site/PageServer?pagename=micro_100landing}}).
Some expert lists, such as the US News ranking of MBA programs, are so
widely accepted and quoted (for better or worse), they have the power
to shape popularity.  

Similarly, it possible that the criteria used to create the expert
rankings are not a good match to the popularity-based hyperlink metrics.
For example, the Fortune 500 list is ranked according to gross sales --
an obvious, impartial metric.  However, we would expect companies such
as Microsoft (Fortune rank \#49) and Target (Fortune rank \#33) to have
a higher SE rank than companies such as Valero Energy (Fortune rank
\#16) and Cardinal Health (Fortune rank \#19) based on the web nature
(Microsoft) and online shopping potential (Target) of their sites.
The WTA and ATP rank their players based on the obvious, quantifiable
metrics of wins and monetary earnings.  However, players' web pages
most likely accrue links based on additional characterics such
as charisma, endorsements and native languge.

On the other hand, it is possible that Money Magazine's choice to feature
only small towns and cities in the 2007 more directly influenced their
page ranking by the SEs.  Prior to their appearance in this expert list,
these places probably had little reason to acquire links and probably
acquired hyperlinks as a mainly as a result of appearing in the Money
expert list.  Even though neither Live nor Yahoo was correlated with
the expert list, they were moderately correlated with each other (and
just missed a moderate correlation at n=50).  This suggests that the
hyperlinks they observed in the web graph were similar and insufficient
for their respective algorithms to arrive at different rankings.  However,
we are at a loss as to why Google had a moderate negative correlation
with the Money expert list.

We also suspect that there might be an optimal age for real-world objects
to drive the popularity of their corresponding web pages.  For example,
the IMDB expert rankings only 9 of the 50 movies were released since 2000,
most are much older.  The \#1 ranked movie according to the IMDB experts
is ``The Godfather'', released in 1972.   While this movie's place in
various expert lists is assured, it is not clear that its corresponding
URLs (either IMDB or Wikipedia) are acquiring hyperlinks at the same
rate as contemporary movies.  This is similar to ``obliteration by
incorporation'' in citation indexing theory: some concepts become
so accepted and pervasive that they no longer require citations
\cite{merton}.  In the other direction, it is possible that fast moving
expert rankings such as Billboard might have real-world objects that
outstrip their web page counterpart's ability to acquire hyperlinks.
For example, the Live API claimed to not have indexed the official
home page of the artist with the number one song according to Billboard,
``Flo Rida'' (although it does show up in the web user interface). 

Finally, it possible that the SE rankings of web pages are significantly
influenced by web-only phenomena that have no correspondence to their
real-world objects.  This could include things such as page update rate,
MIME type, \url{robots.txt} files, etc.

\section{Conclusions and Future Work}

Inspired by the question of Amento et al. \cite{amento:auth-qual}
``Does Authority mean Quality?'', we have asked ``Does Quality mean
Authority?''  We tested 57 expert-SE rankings and 42 inter-SE rankings.
Of those 99 tests, only 11 had statistically significant moderate or
high correlations.  No SE stands out as more correlated than the others:
Yahoo was correlated with Live 4 times, Live with Google once and Google
with Yahoo once.  Live, Google and Yahoo correlated with the experts
once, twice and twice respectively.  Mapping from real-world objects
to corresponding web pages is difficult and this may have contributed
to the low number of correlations.  However, we were surprised to not
find more inter-SE correlation.  Although we cannot say we disproved
correlation between expert rankings of real-world objects and the search
engine rankings of their corresponding web pages, we have shown there
are numerous scenarios where we believe it is reasonable to expect
a correlation (especially in lists (e.g., IMDB, Billboard) where
quality is a function of popularity), but this correlation is absent.
Although highly ranked pages are likely to be considered quality pages
by experts, we cannot be sure that real-world resources (e.g., athletes,
movies, universities) considered to be quality by experts will have
highly ranked corresponding web pages.  To answer our question, although
authority means quality, quality does not necessarily mean authority.

We consider these results to be baseline for future research.  There
are obvious areas to improve and extend the methodology and tests
presented here.  First, more and different expert lists could be used.
For a particular topic, differing experts could be used (e.g., AFI
vs. IMDB), and more topics could be explored (sports teams in addition to
individual athletes, contemporary movies, consumer product reviews, etc.).
In particular, more research should be done in profiling the optimum age
for a real-world object / web page pair.  We expect that as age increases,
the ranking of the real-world object will continue to climb or hold its
value, while the ranking of the web pages will likely give ground to newer
web pages.  The English language / United States bias should be removed.

Our immediate next step in our research is to expand the limitation of
a single URL per real-world object.  While this is less of a limitation
for universities and businesses, artists and athletes are likely to have
multiple candidate URLs.  We are working on a refinement to our program
to query a SE for a real-world object (e.g., ``Roger Federer'') and
record the top $k$ resulting URLs.  These URLs would then be aggregated
to more accurately calculate the rank of a real-world object in a SE.
This is a problem for all hyperlink derived metrics: multiple candidate
URLs can compete for a limited number of links on web pages, thereby
reducing their importance or popularity metric.

\bibliographystyle{abbrv}
\bibliography{mln} 

\end{document}